\documentclass[twocolumn]{aastex62}

\usepackage{graphicx}
\usepackage{booktabs}
\shorttitle{CNN BAL Classification}
\shortauthors{Guo \& Martini}
\graphicspath{{./}{figure/}}

\newcommand{\kms}{{\rm{\,km\,s^{-1}}}}

\begin{document}

\title{Classification of Broad Absorption Line Quasars with a Convolutional Neural Network} 

\author{Zhiyuan Guo}
\affiliation{Department of Astronomy, The Ohio State University,
140 West 18th Avenue, Columbus, OH 43210, USA\\
guo.920@osu.edu}

\author{Paul Martini}
\affiliation{Department of Astronomy and Center for Cosmology and Astro-Particle Physics, 
The Ohio State University, Columbus, OH 43210, USA\\
martini.10@osu.edu}

\begin{abstract}
Quasars that exhibit blue-shifted, broad absorption lines (BAL QSOs) are an important probe of black hole feedback on galaxy evolution. Yet the presence of BALs is also a complication for large, spectroscopic surveys that use quasars as cosmological probes because the BAL features can affect redshift measurements and contaminate information about the matter distribution in the Lyman-$\alpha$ forest. We present a new BAL QSO catalog for quasars in the Sloan Digital Sky Survey (SDSS) Data Release 14 (DR14). As the SDSS DR14 quasar catalog has over 500,000 quasars, we have developed an automated BAL classifier with a Convolutional Neural Network (CNN). We trained our CNN classifier on the \ion{C}{4} $\lambda 1549$ region of a sample of quasars with reliable human classifications, and compared the results to both a dedicated test sample and visual classifications from the earlier SDSS DR12 quasar catalog. Our CNN classifier correctly classifies over 98\% of the BAL quasars in the DR12 catalog, which demonstrates comparable reliability to human classification. The disagreements are generally for quasars with lower signal-to-noise ratio spectra and/or weaker BAL features. Our new catalog includes the probability that each quasar is a BAL, the strength, blueshifts and velocity widths of the troughs, and similar information for any \ion{Si}{4} $\lambda 1398$ BAL troughs that may be present. We find significant BAL features in 16.8\% of all quasars with $1.57 < z < 5.56$ in the SDSS DR14 quasar catalog. 

\end{abstract}

    
\section{Introduction} \label{sec:intro}

Quasars or quasi-stellar objects (QSOs) are highly energetic sources at the centers of galaxies that are caused by the accretion of matter onto supermassive black holes. A sub-set of QSOs exhibit blue-shifted, broad absorption line (BAL) troughs with velocities greater than $2000 \kms$ \citet{Weymann91}. One longstanding question in BAL research is if the BAL phenomenon represents an evolutionary phase of all QSOs, or if it is always present, but only visible along a subset of all lines of sight. One interesting observation is that the fraction of BALs in a QSO sample depends on the selection method. Surveys that employ UV and visible wavelengths find BAL fractions of 10-30\% \citep{Foltz90,Trump06}, while the IR-selected BAL fraction is greater and about 40\% based on a 2MASS-selected sample from \citet{Dai08}. This result suggests that BALs are present in at least a large fraction of all QSOs, and that the presence of BAL troughs may inhibit the identification of BAL QSOs via UV and visible-wavelength selection methods. 

BAL QSOs are often subdivided based on the spectral lines that show broad absorption features, and the BAL fraction depends on which absorption features are seen. The most common type of BAL QSO just exhibits absorption in high-ionization lines such as \ion{C}{4} $\lambda 1549$ and are called HiBALs. If absorption troughs from low-ionization features such as \ion{Mg}{2} are also seen, then the BAL is classified as a LoBAL. LoBALs that exhibit absorption in Fe lines such as \ion{Fe}{2} are classified as FeLoBALs. Finally, the rarest BALs exhibit absorption in the Balmer lines \citep{Hall07,Mudd17}. The distribution of these classes depends on selection method, as \citet{Trump06} find HiBALs, LoBALs, and FeLoBALs are 26\%, 1.3\%, and 0.3\% in their study of QSOs in the SDSS, while \citet{Urrutia09} find that the BAL fraction is above 30\% for all three classes based on an IR-selected sample. 

One important application of large, spectroscopic quasar samples is to measure the matter distribution at $0.8 < z < 2.2$, the redshift range where quasars are the most accessible tracer of the matter distribution \citep{Ata18}. That analysis requires robust measurements of quasar redshifts and their corresponding uncertainties. For typical quasars, redshifts may be calculated by visual inspection, from one or more prominent emission lines, or principal component fits to the entire observed spectrum. Many studies have investigated the best methods to estimate redshifts and quantify the uncertainties \citep[e.g.][]{Hewett10,Paris12}. The uncertainties in these procedures are greater if the quasar is a BAL.

Starting at redshifts of about $z > 2$, quasars may also be used to measure the Baryon Acoustic Oscillation (BAO) signal in the Ly$\alpha$ forest \citep{McDonald07}, and the first detection was reported by \citet{Font14}. These studies typically exclude BAL QSOs \citep[see also][]{Bautista17} because the BAL features can be present in the Ly$\alpha$ forest, where they cannot be readily distinguished from the Ly$\alpha$ absorption from intervening clouds of neutral gas. While \ion{C}{4} absorption does not affect the rest-frame $1040 < \lambda < 1200$\AA\ region commonly used for cosmology studies, the presence of collisionally-ionized \ion{C}{4} is a good indication that other strong, UV absorption features such as \ion{N}{5} $\lambda\lambda 1239, 1243$ and \ion{O}{6} $\lambda\lambda 1032, 1038$ will be present, and these lines can impact the forest region for a range of blueshifts. Other lines are sometimes detected in BALs as well, including \ion{P}{5} $\lambda\lambda 1118, 1128$ and Ly$\alpha$ \citep{FilizAk14,Hamann19}. 

Identification of BAL quasars in large spectroscopic samples has become a progressively greater challenge as the sample sizes have increased. \citet{Paris17} performed a visual classification of all QSOs through the SDSS Data Release 12 (DR12) quasar catalog, which had 297,301 quasars. This includes 29,580 BAL quasars. However, the DR14 sample of 526,356 quasars (this includes previous releases) is only partially based on visual inspection and classification. This catalog identifies 21,877 BAL quasars, and it does not provide the same level of information about BAL features as previous catalogs, such as trough blueshifts and widths. Spectroscopic datasets from the upcoming Dark Energy Spectroscopic Instrument \citep{Desi16a,Desi16b} and 4MOST \citep{deJong14} will be even larger than these current generation surveys. 

The visual identification of BAL features in progressively larger QSO samples has become progressively more time consuming. In addition, the subjective aspects of human classification, especially if split amongst several humans, add additional complications to statistical studies. A good alternative is to use machine learning, as this approach can provide both uniform classification and a robust measurement of reliability. A Convolutional Neural Network (CNN) is particularly well suited to this application, as this type of artificial neural network is designed to mimic the behavior of the visual cortex. This type of network has already shown great promise in application to spectroscopic datasets. For example, \citet{Parks18} used a CNN to identify and quantify the properties of Damped Lyman-$\alpha$ systems in SDSS quasars at a broad range of redshifts. In another recent study, \cite{Busca18} introduced  QuasarNET, which uses a CNN to estimate redshifts and classify quasar candidates as either stars, galaxies, or quasars, and identified BALs in the quasar sample. Their study applied QuasarNET to the SDSS-DR12 sample of \citet{Paris17}. 

In this paper, we develop and apply a CNN to the classification of BAL QSOs in the SDSS DR14 Quasar Catalog of \citet{Paris18}, and provide a catalog of the BAL properties of these quasars. In \S~\ref{sec:method} we provide the basic quantities we calculate for BALs and give more details of the datasets in this analysis. In \S~\ref{sec:ML} we describe the development, training, and testing of our CNN classifier. Then in \S~\ref{sec:compare} we apply our classifier to the DR14 quasar catalog and analyze its performance relative to other catalog data. We summarize our results in the last section.  

\section{Methodology} 
\label{sec:method}

\subsection{BAL Characterization} \label{sec: BAL Identification}

BAL quasars are often described by two metrics that quantify their \ion{C}{4} absorption troughs: the balnicity index BI proposed by \citet{Weymann91} and the intrinsic absorption index AI proposed by \citet{Hall02}. BI is computed from $25,000$ to $3000 \kms$ blueshift relative to \ion{C}{4}:  
\begin{equation}\label{eq:bi}
BI  = -\int_{25000}^{3000}\left[1 - \frac{f(v)}{0.9}\right]C(v)dv
\end{equation}
The quantity $f(v)$ is the normalized quasar flux density as a function of velocity displacement from the \ion{C}{4} emission line. It is computed relative to an estimate of the intrinsic continuum of the quasar. The function $C(v)$ is in a binary function that is set to one once the quantity $[1-\frac{f(v)}{0.9}]$ is continuously positive for more than $2000 \kms$ and set to zero otherwise. The error on balnicity index is given by \citet{Trump06} as:
\begin{equation}\label{eq:sigma_bi}
\sigma_{BI}^{2} = -\int_{25000}^{3000}\left(\frac{\sigma_{f(v)}}{0.9}\right)^{2}C(v)dv
\end{equation}
The quantity $\sigma_{f(v)}$ is the uncertainty on the flux in each pixel of the normalized flux density. One drawback of this approach is that it assumes perfect knowledge of the quasar continuum shape, which can be especially uncertain on the blue wing of the \ion{C}{4} line. Another is that it is not sensitive to BAL troughs that are very shallow. We attempt to address the uncertainty due to the continuum with the addition of an error term that accounts for the uncertainty due to the continuum fit. Our modified equation is: 
\begin{equation}
\sigma_{BI}^{2} = -\int_{25000}^{3000}\left(\frac{\sigma_{f(v)}^{2}+\sigma_{PCA}^{2}}{0.9^2}\right)C(v)dv
\end{equation}
The quantity $\sigma_{PCA}$ is the uncertainty in our PCA fitting. We describe how we calculate this quantity in Section~\ref{sec:pca}. 
AI and the uncertainty in AI are similar to the expressions for BI:
\begin{equation}
AI = -\int_{25000}^{0}\left[1 - \frac{f(v)}{0.9}\right]C(v)dv\\
\end{equation}
\begin{equation}
\sigma_{AI}^{2} = -\int_{25000}^{0}\left(\frac{\sigma_{f(v)}^{2}+\sigma_{PCA}^{2}}{0.9^2}\right)C(v)dv
\end{equation}
This main difference between AI and BI is that for AI the quantity $C(v)$ is set to one once the trough has extended for more than $450 \kms$, rather than $2000 \kms$. It also extends the calculation velocity interval to zero blueshift. \citet{Hall02} introduced AI to account for uncertainties in the systemic redshift, the continuum shape, and to measure intrinsic absorption systems, such as the mini-BAL troughs identified by \citet{Hamann01}. 

Finally, \citet{Trump06} introduced the parameter $\chi^2_{trough}$, the reduced chi-squared for each detected trough:
\begin{equation}
	\chi^2_{trough} = \sum\frac{1}{N}\left[\frac{1-f(v)}{\sigma}\right]^2
\end{equation}
In this expression $N$ is the number of pixels in a trough, $f(v)$ is the normalized flux, and $\sigma$ is the estimated rms noise for each pixel. This expression is intended to quantify the statistical significance of any apparent trough, such that larger values correspond to more significant troughs. This quantity is particularly useful for assessing weak troughs and/or troughs in low signal-to-noise ratio data. \cite{Trump06} consider a trough with $\chi^2_{trough} > 10$ to be significant.

\subsection{Data} \label{sec:data}

The starting point for our analysis is the SDSS DR14 Quasar Catalog of \citet{Paris18}. This catalog has 526,356 quasars, including measurements of BI and $\sigma_{BI}$ for each quasar. The catalog was derived from the spectroscopic data in the Fourteenth Data Release of SDSS \citep{Abolfathi18}, and we also used these spectra as the starting point for our analysis. These spectra were mostly obtained as part of the Baryon Oscillation Spectroscopic Survey (BOSS) and its extension \citep[eBOSS][]{Dawson13,Dawson16}, which were observed with the SDSS spectrograph \citep{Smee13}. More information about the selection and analysis of these quasars are described in \citet{Paris18}. 

We also used information from the SDSS DR12 quasar catalog \citep{Paris17}. That catalog includes 297,301 quasars from the Twelfth Data Release of SDSS \citep{Alam15}. The advantage of the DR12 catalog is that BAL quasars were flagged during a visual inspection of all of the quasar targets \citep[see also][]{Paris12}. The DR12 quasar catalog has 29,580 quasars visually flagged as BALs. For all quasars at $z \geq 1.57$, this catalog includes measurements of BI, AI, and $\chi^2_{trough}$. There is an AI value if there is at least one trough with $\chi^2_{trough} \geq 10$, and 48,863 quasars meet this criterion. The catalog also includes the number of troughs and the velocity ranges of each trough. Of the  sample of 29,580 quasars visually flagged as BALs, 21,444 have AI$ > 0$ and 15,044 have BI$ > 0$. 

\begin{figure}[t] 
\centering
\includegraphics[scale = 0.45]{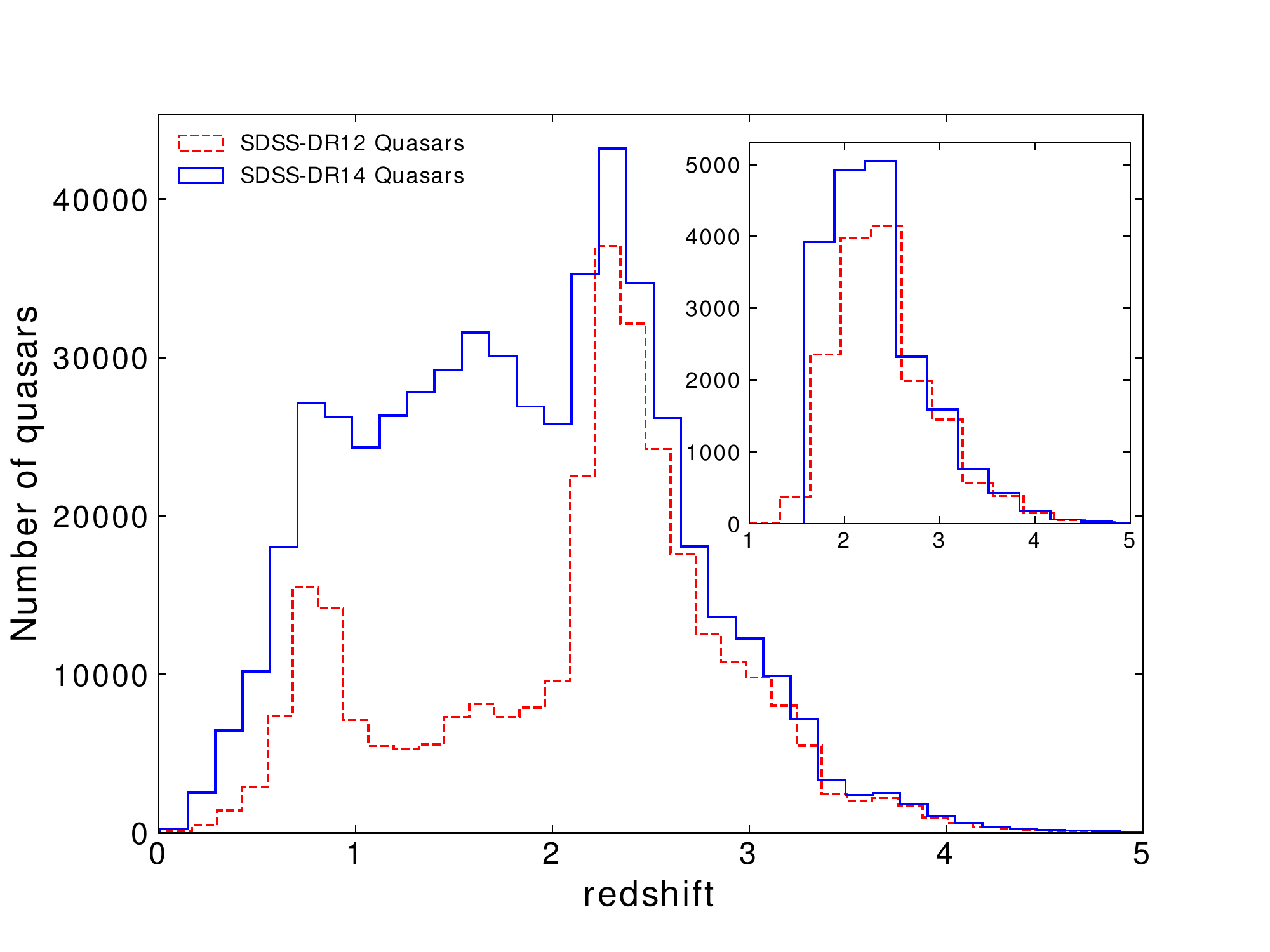}
\caption{Redshift distribution of the SDSS-DR12 ({\it dashed red histogram}) and SDSS-DR14 ({\it solid blue histogram}) quasars over the redshift range $0 < z < 5$. The inset panel in the upper right shows the redshift distribution of quasars with a significant BI value, $BI > 3\sigma_{BI}$. The BAL quasars are only shown over the redshift range $1.57 < z < 5.56$ where the \ion{C}{4} region is visible with the SDSS spectra.} 
\label{fig:dr12_14_z_compare} 
\end{figure}

The redshift distributions of the DR12 and DR14 quasar catalogs are shown in Figure~\ref{fig:dr12_14_z_compare}. The DR14 catalog includes many more quasars with $0.8 < z < 2.2$ because of a change in the selection criteria to identify more quasars to trace large-scale structure in this redshift range \citep{Ata18}. The inset panel shows the redshift distribution for quasars with a significant BI value, which we define as $BI > 3\sigma_{BI}$. 

\subsection{Principal Component Analysis} \label{sec:pca}

\begin{figure}[t]
\includegraphics[scale = 0.45]{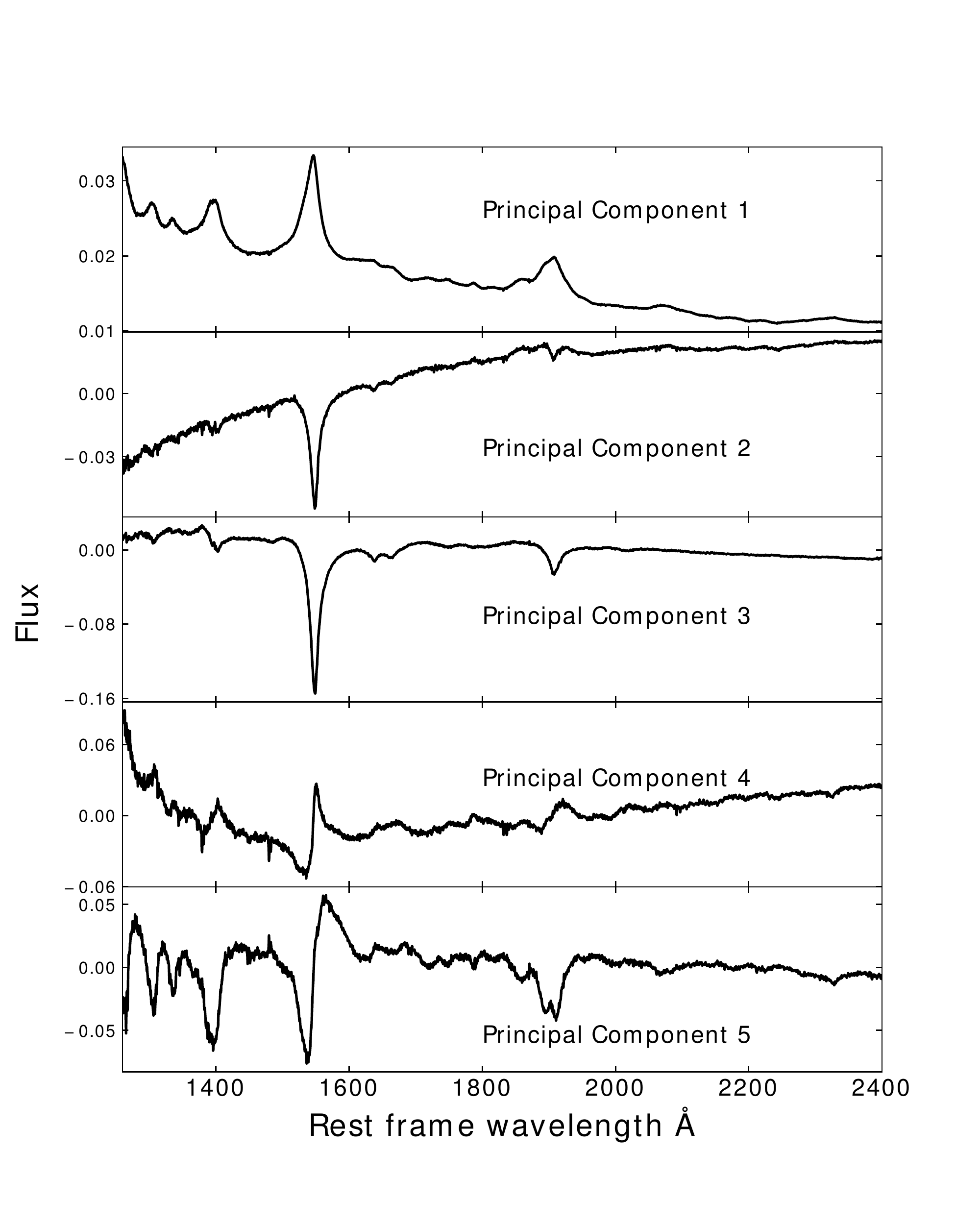}
\caption{Five principal components computed from a sample of 8000 quasars with no absorption features. The PCA components span the rest frame wavelength range 1260\AA\ to 2400\AA. See Section~\ref{sec:pca} for more details.}
\label{fig:5pca}
\end{figure}

We use Principal Component Analysis (PCA) to fit the spectra of the quasars from DR14. These fits are necessary to obtain accurate estimates of the continuum to characterize any absorption troughs. Removal of the quasar continuum shape and broad emission features also reduces the complexity of the quasar spectra for automated classification. Similar to \citet{Paris12}, we generated five principal components from 8000 quasars with no evidence for BAL features, redshifts of $1.57 < z < 5.56$ that match our search for BALs, and generate PCA components over the rest frame wavelength range from 1260\AA\ to 2400\AA. This wavelength range provides good coverage of the \ion{C}{4} and \ion{Si}{4} regions where we characterize absorption troughs with blueshifts up to $25000 \kms$. The five PCA components are shown in Figure~\ref{fig:5pca}.

\begin{figure*}[t]
\includegraphics[scale = 0.5]{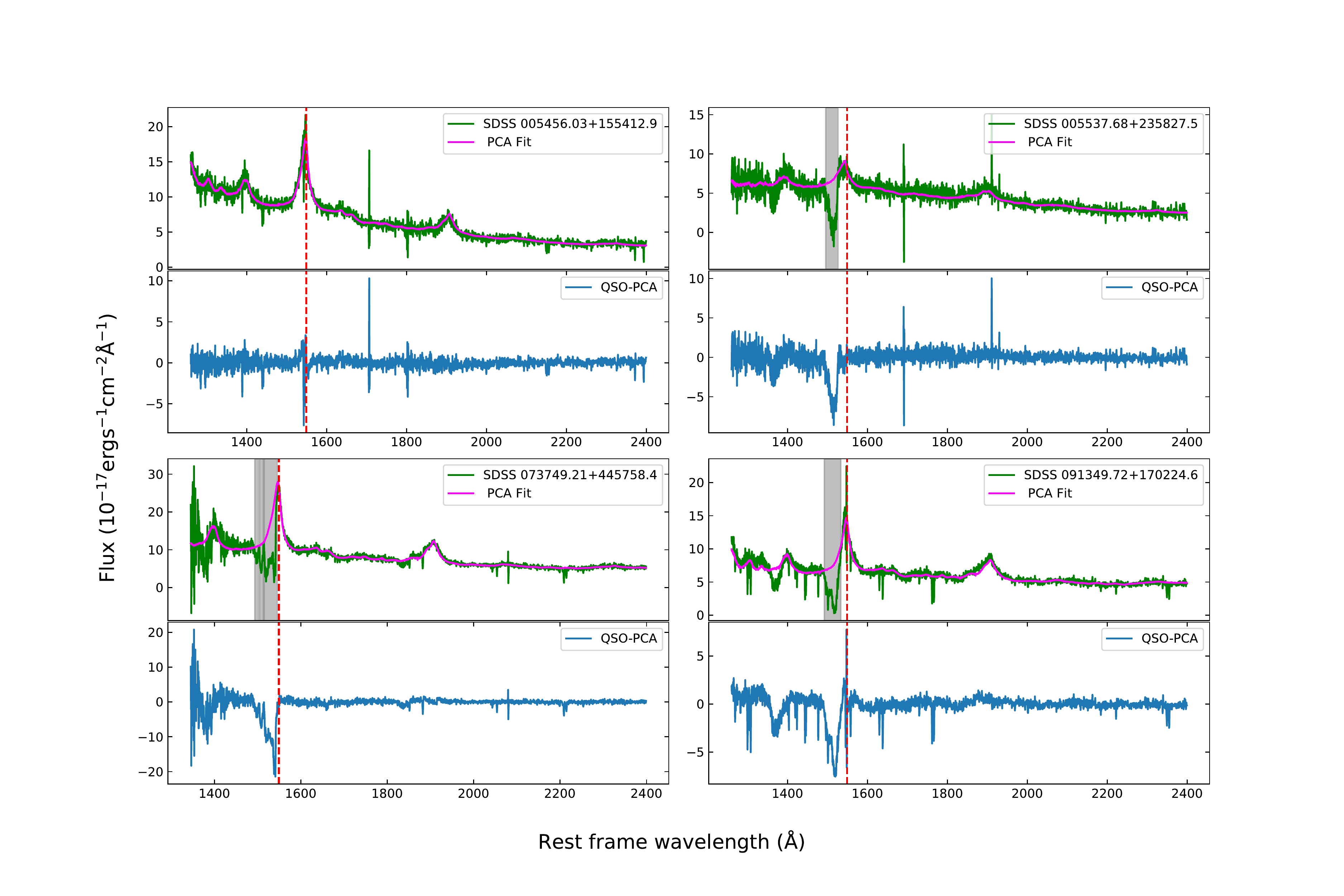}
\caption{Example PCA fits for a non-BAL quasar ({\it upper left}) and three BAL quasars ({remaining panels}). Each panel shows the wavelength range from 1260\AA\ to 2400\AA, including the \ion{C}{4} line ({\it vertical, dashed line}). Each panel has two parts. The upper part shows the original spectrum and the PCA fit, while the lower part show the difference of the quasar spectrum and the PCA fit. Any significant BAL feature associated with \ion{C}{4} is masked during the PCA fit ({\it vertical, shaded regions}). }
\label{fig:pca_examples}
\end{figure*}

We fit these five PCA components to each quasar with a ${\chi}^2$ minimization algorithm. This algorithm decreases the wavelength range of the fit for quasars near the redshift limits of our study. We also run an algorithm to detect troughs with blueshifts from $-25000$ to $0 \kms$ relative to \ion{C}{4}, and use the results to iteratively mask BAL features. The iterative masking of the BAL features significantly improves the PCA fit to quasars with significant absorption troughs. Finally, we subtract the best PCA fit from each quasar. Examples of the PCA fit and the subtraction are shown in Figure~\ref{fig:pca_examples}.

We calculated PCA fits to all of the SDSS DR14 quasar spectra with $1.57 < z < 5.56$. In most cases, the subtracted spectrum is flat and the broad emission lines of \ion{C}{4} and other ions are barely visible, especially for quasars without BAL troughs. The exceptions are typically on the blue side of the \ion{C}{4} line (and often the \ion{Si}{4} line), where the impact of absorption troughs are most apparent. These difference spectra therefore highlight exactly the features that we want the automated classifier to identify. Finally, we only use the velocity range from $-25000$ to $0 \kms$ relative to \ion{C}{4} for the automatic classification. This dramatically decreases the size of the data, and increases the efficiency of the classifier. 

Following the method described above, we fit PCA components to all of the quasars in SDSS DR14. Figure \ref{fig:pca_examples} shows examples for both non-BAL and BAL quasars centered on the \ion{C}{4} emission line. This emission line is removed when the PCA fit is subtracted, while the absorption features in the BAL quasar examples are preserved. Even though we fit the PCA components over a wide wavelength range, we only use the subtracted spectra from $-25000 \kms$ to $0 \kms$ relative to \ion{C}{4} as input to our classifier. 

\begin{figure}
\includegraphics[scale = 0.45]{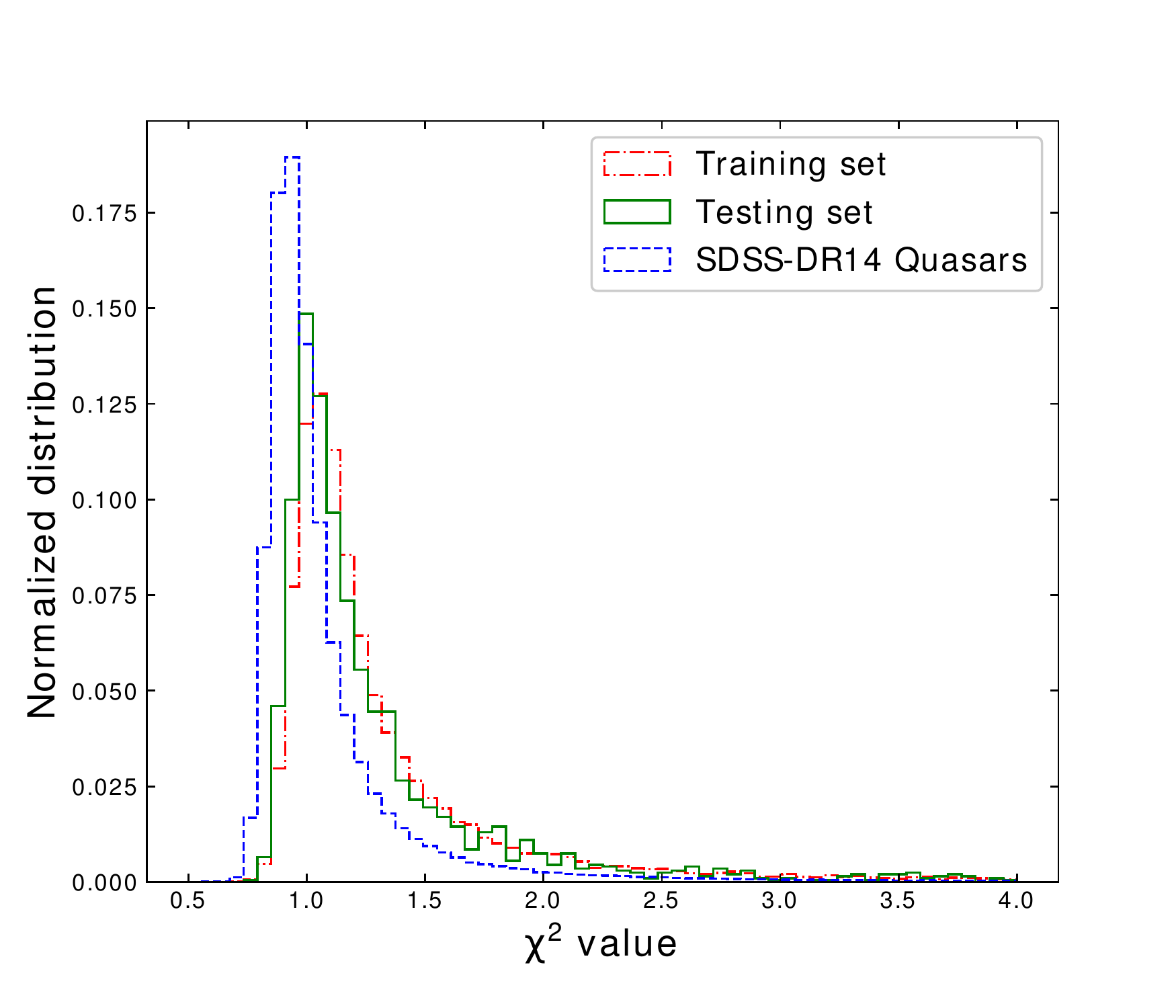}
\caption{Reduced ${\chi}^2$ distribution of the PCA fits to the entire sample of SDSS-DR14 quasars ({\it blue, dashed line}). The distributions are also shown for the CNN training set ({\it red, dotted line}) and test set ({\it green, solid line}). These two subsets of the data have larger ${\chi}^2$ values because both have a larger percentage of BALs by construction (about 50\%), and the fits to BALs typically have larger ${\chi}^2$ values.}
\label{fig:pcadist}
\end{figure}

To characterize the quality of the PCA fit for the measurement of absorption features, we calculate a separate ${\chi}^2_{fit}$ over just the velocity range from $-25,000$ to $5000 \kms$ relative to \ion{C}{4}: 
\begin{equation}
{\chi}^2_{fit} = \frac{1}{D}\sum_{k=1}^n \frac{(O_k - E_k)^2}{\sigma^2}
\end{equation}
The quantity $D$ is the number of degrees of freedom, $O_k$ and $E_k$ represent the value for each pixel in the original spectrum and the PCA fit, respectively, and $\sigma$ is the uncertainty in each original pixel as provided in \citet{Paris17}. The distribution of these ${\chi}^2_{fit}$ values for our DR14 quasar sample are shown in Figure~\ref{fig:pcadist}. The distribution of ${\chi}^2_{fit}$ values is peaked at about one, although with a tail to larger values.

\section{Automatic BAL Classification} \label{sec:ML}

We chose to implement our automatic BAL classification algorithm with a convolutional neural network (CNN), a deep learning method that is particularly well suited to applications that are analogous to visual imagery, such as image classification. CNN classifiers also provide a probability that a given object is a member of a class, a quantization of the confidence that a human classifier may have, which has many advantages for statistical analyses of the data. Other options, such as supervised learning methods and probabilistic classifiers, generally work best when the subject of the classification can be described by a number of distinct properties. The wide range of blueshifts and trough shapes exhibited by BALs are not that amenable to the feature engineering required of those other options.

\subsection{CNN Structure}\label{sec:cnn}

\begin{figure}[t]

\centering
\includegraphics[scale = 0.3]{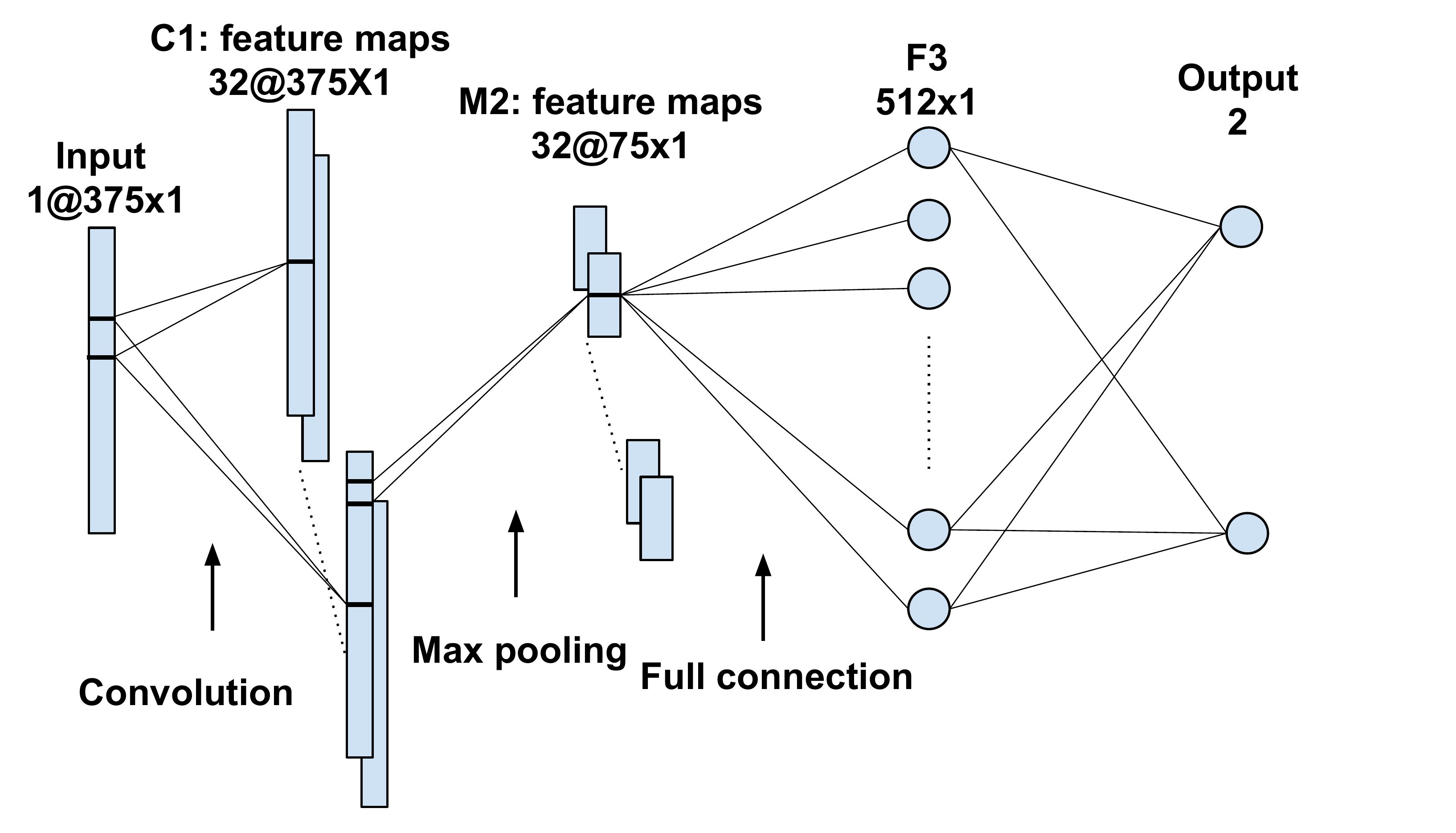}

\caption{CNN structure employed to create our BAL classifier. Each input spectrum has 375 pixels. The CNN structure has one convolutional layer, one max pooling layer, and a fully connected layer that performs the classification. See Section~\ref{sec:ML} for more details.}
\label{fig:cnn}
\end{figure}

A typical CNN structure contains layers of several types, in addition to the input and output layers. The three types of layers in our application are a convolutional layer, which applies operations on the input to create feature maps, a pooling layer, which downsamples the information from the feature maps to better identify spatial hierarchies, and a dense layer or fully connected layer, which performs the classification on features extracted from the preceding layers. The output is the probability that a quasar is a BAL. We experimented with varying the number of layers and convolution kernels with a range of sizes and found we achieved good results with a relatively shallow structure: a single, one-dimensional convolutional layer with 32 convolution filters, a single, one-dimensional pooling layer with max pooling that downsamples the data by a factor of five, and a fully connected layer. Two virtues of this simple approach are that the relatively modest number of parameters avoids overfitting the training set and the computation times are very modest. This CNN structure works well because of the relatively small size of our dataset (375 pixels per object) and the relatively simple nature of the classification problem after subtraction of the PCA fit. As CNNs typically work best with input layers on the interval [0,1] or [-1,1], we also experimented with various schemes to renormalize our data, such as division by the PCA fit, but did not identify one that produced substantially better results than simple subtraction. The CNN structure we use is shown in Figure~\ref{fig:cnn}. We implemented the CNN structure with TensorFlow\footnote{https://www.tensorflow.org}, an open source software library for machine learning \citep{Abadi15}.

\subsection{Training and Testing Sets} \label{sec:train}

Machine learning methods such as our CNN classifier require a training set. We used the DR12 quasar catalog of \citet{Paris17} as the starting point to produce one for our classifier. One virtue of the DR12 catalog is that there was a visual inspection of every quasar. However, we did not completely rely on the DR12 classifications because human classification is inherently subjective and there is no single, quantitative definition of a BAL that is appropriate for all applications. For example, the balnicity index of \citet{Weymann91} does not include the first $2000 \kms$ of the absorption feature, will miss shallow troughs, and does not extend to the center of the \ion{C}{4} line. It will consequently miss broad absorption that is less than $2000 \kms$ in extent that could nevertheless impact cosmological analysis with the Ly$\alpha$ forest, as well as strong absorption features near the center of the \ion{C}{4} line that could compromise the redshift estimate. The AI quantity introduced by \citet{Hall02} is sensitive to narrower absorption features that are still broader than typical galaxy velocities, and does extend to zero velocity, although it is still insensitive to the shallowest features. Finally, both of these measures work less well in low signal-to-noise ratio spectra, and both can be compromised by poor fits to the quasar continuum. 

We started construction of our training set with about 10,000 visually-classified BALs and 10,000 visually-classified non-BALs from the DR12 quasar catalog \citep{Paris17}, but adjusted these classifications through several iterative passes. For each iteration we trained a new classifier on the BAL and non-BALs in the training set, ran the classifier on the training sample, visually inspected all of the apparent mis-classifications and ambiguous cases, and adjusted the classifications of the training set as appropriate. After several iterations, the classifier converged well with our visual classifications. In all cases, we label quasars with a BAL probability greater than 50\% as BALs, and quasars below this percentage as non-BALs. For our visual inspection step, we classified a quasar as a BAL if it showed narrower troughs than the $2000 \kms$ minimum width for BI, and included troughs that extended to the center of the \ion{C}{4} emission line. Our iterative process changed the final classifications of about 6.5\% of the training set quasars relative to the DR12 visual BAL flag. While still a subjective process, this approach proved to be an efficient way to build a substantive training sample whose visual classification criteria are known to us, and should include the full range of absorption troughs that could impact cosmological analysis. We then used this training set to construct a new classifier that we applied to the test set and the entire DR14 quasar catalog. 

We emphasize that our visual classification criteria were chosen to identify any evidence of absorption features that could impact cosmological analysis, and consequently our definition of BAL will include quasars that do not meet the criteria introduced by \citet{Weymann91}, nor even have the intrinsic absorption index introduced by \citet{Hall02}. For example, only 98.9\% of the quasars classified as BALs in our training set have significant AI measurements, and only 62.5\% have significant BI measurements. In contrast, 24.6\% of the non-BAL quasars in our training set have significant AI measurements, and 3.3\% have significant BI measurements. This demonstrates that while we classify more quasars as BALs than we would have by some other definitions (e.g. use of BI), we are not missing significant numbers of BALs.  

We developed a test set of 2000 other quasars to test the performance of the classifier. This test set was selected to have approximately equal numbers of BAL and non-BAL quasars based on the DR12 visual classifications, although we adjusted the classifications in some cases to align with the criteria we used for the training set. We then classified all of these quasars with the CNN classifier and  visually inspected the output.  We found that we agreed with the classifier in nearly all cases. The BAL classifier also showed similar agreement with the DR12 visual classification as it did with the test set, as 86.0\% of the BALs identified by our CNN classifier have a visual BAL classification from DR12. And similar to the test set, 98.6\% of the BALs identified by the classifier have significant AI measurements and 61.1\% have significant BI measurements. This sample is also not missing significant numbers of BALs. While 7.9\% of the non-BAL quasars in the test set are visually classified as BALs in the DR12 catalog, only 7.5\% and 2.6\% have significant AI and BI measurements, respectively. Finally, we visually inspected the ambiguous cases (probabilities near 50\%) and found that most of them either do not have clear BAL features (the absorption features are either too shallow or too narrow) or the spectra have sufficiently low SNR that  absorption features are unclear.

\subsection{DR14 BAL catalog} \label{sec:dr14balcat}

\begin{figure}[t]
\centering
\includegraphics[scale = 0.4]{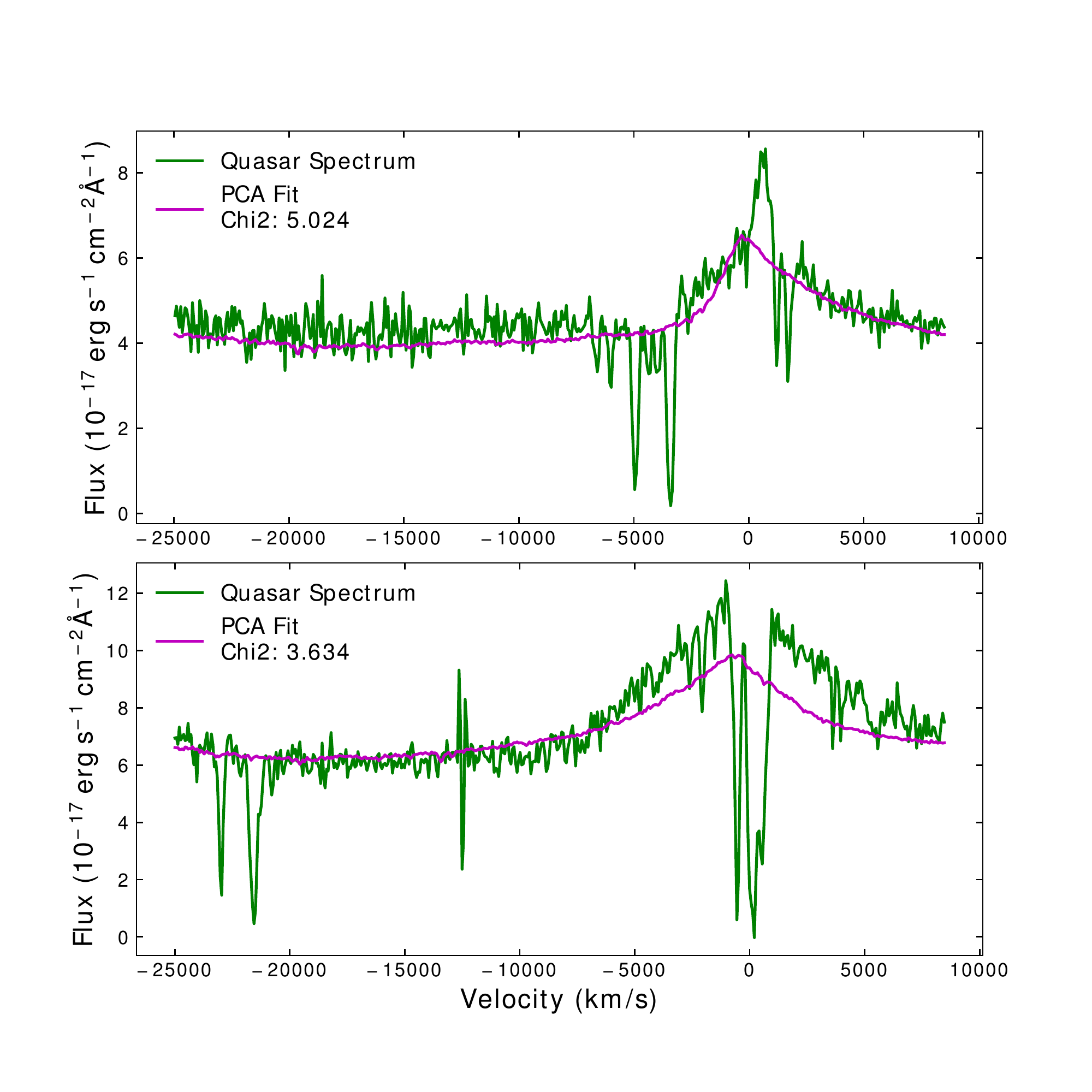}
\caption{Two examples of PCA fits with ${\chi}^2_{fit} > 2$.  {\it Top Panel:} Example of a fit with ${\chi}^2_{fit} = 5.02$. This fit is poor because of the unusual, asymmetric shape of the \ion{C}{4} line. {\it Bottom Panel:} Example of a fit with ${\chi}^2_{fit} = 3.63$. The fit is significantly compromised by the presence of pronounced absorption at the center of the \ion{C}{4}. Not enough of the absorption feature is blueshifted that the absorption was masked by our iterative PCA fit algorithm. }
\label{fig:bigchi2}
\end{figure}

There are 320,821 quasars in the DR14 Quasar Catalog in the redshift range $1.57 < z < 5.6$ where \ion{C}{4} BAL features could be detected. We fit these quasars with the PCA components shown in Figure~\ref{fig:5pca} with the iterative masking procedure described in Section~\ref{sec:pca}, subtracted the PCA fits, and applied the CNN classifier to the velocity range from $-25,000$ to $0 \kms$ of \ion{C}{4}. Figure~\ref{fig:pcadist} shows that the reduced ${\chi}^2_{fit}$ distribution for the DR14 sample is peaked at about one, although with a tail of larger ${\chi}^2_{fit}$ values. The figure also shows the ${\chi}^2_{fit}$ distribution for the training and test sets. Both of these distributions peak at larger ${\chi}^2_{fit}$ values than the full DR14 quasar sample. This is because the quality of the PCA fits to the BAL quasars are generally poorer, even after the troughs are masked, and BALs are overrepresented in the training and test sets. We visually inspected a representative sample of poor fits and generally find the worst agreement around the \ion{C}{4} emission line. Figure~\ref{fig:bigchi2} shows two examples of PCA fits with ${\chi}^2_{fit} > 2$.

\begin{figure}[t] 
\includegraphics[scale = 0.45]{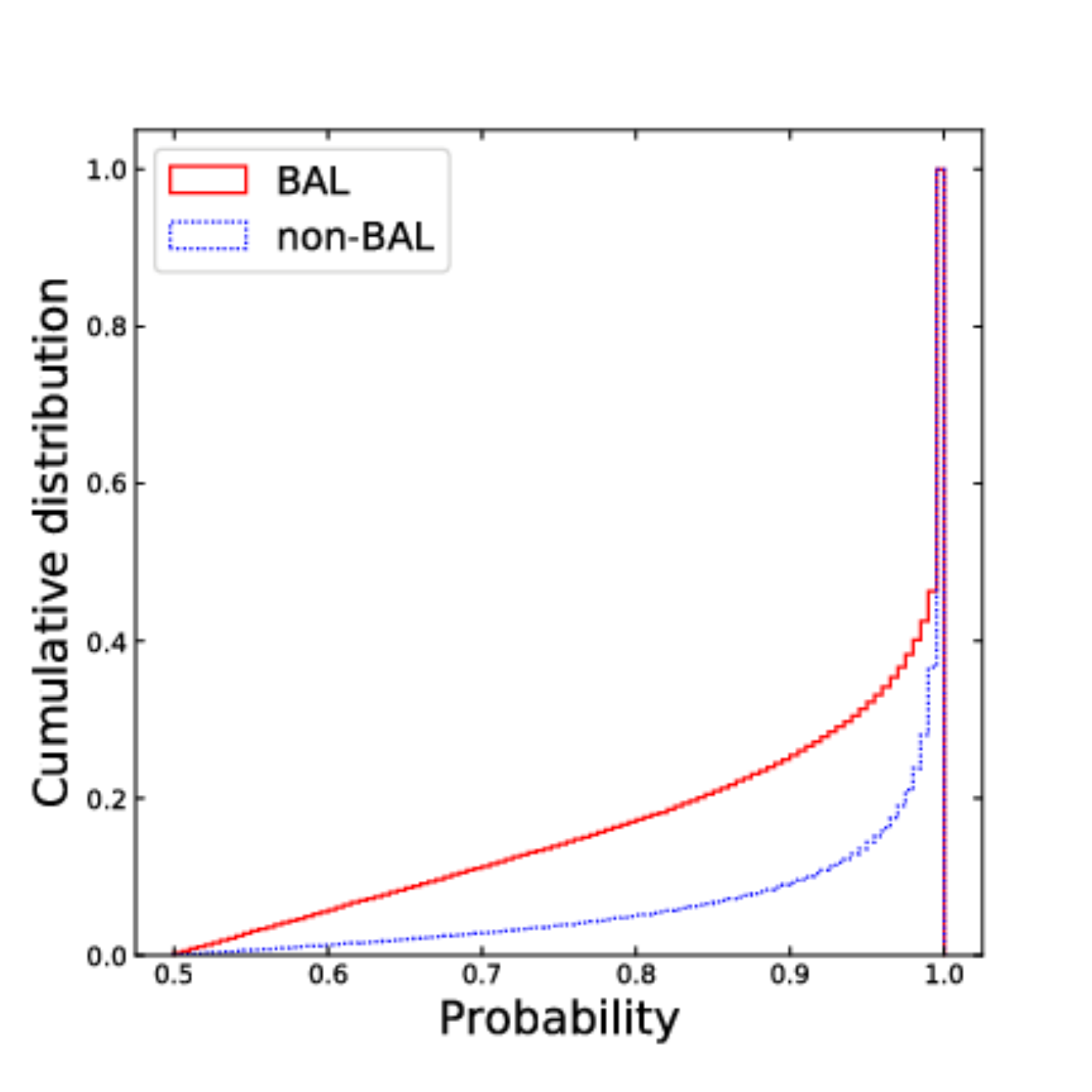}
\centering
\caption{Cumulative probability distribution of DR14 non-BAL quasars ({\it dashed blue}) and BAL quasars ({\it solid red}). Both distributions start at a probability of 0.5, as the probability a quasar is a BAL is $P_{{\rm BAL}} = 1 - P_{{\rm non-BAL}}$. The BAL probability $P_{{\rm BAL}}$ does not approach unity as quickly as  $P_{{\rm non-BAL}}$, which indicates the classifier is generally more confident of a non-BAL classification than a BAL classification. See Section~\ref{sec:dr14balcat} for more details.}
\label{fig:prob_dist}
\end{figure}

\begin{figure*}[p]
\centering
\epsscale{1.2}
\plotone{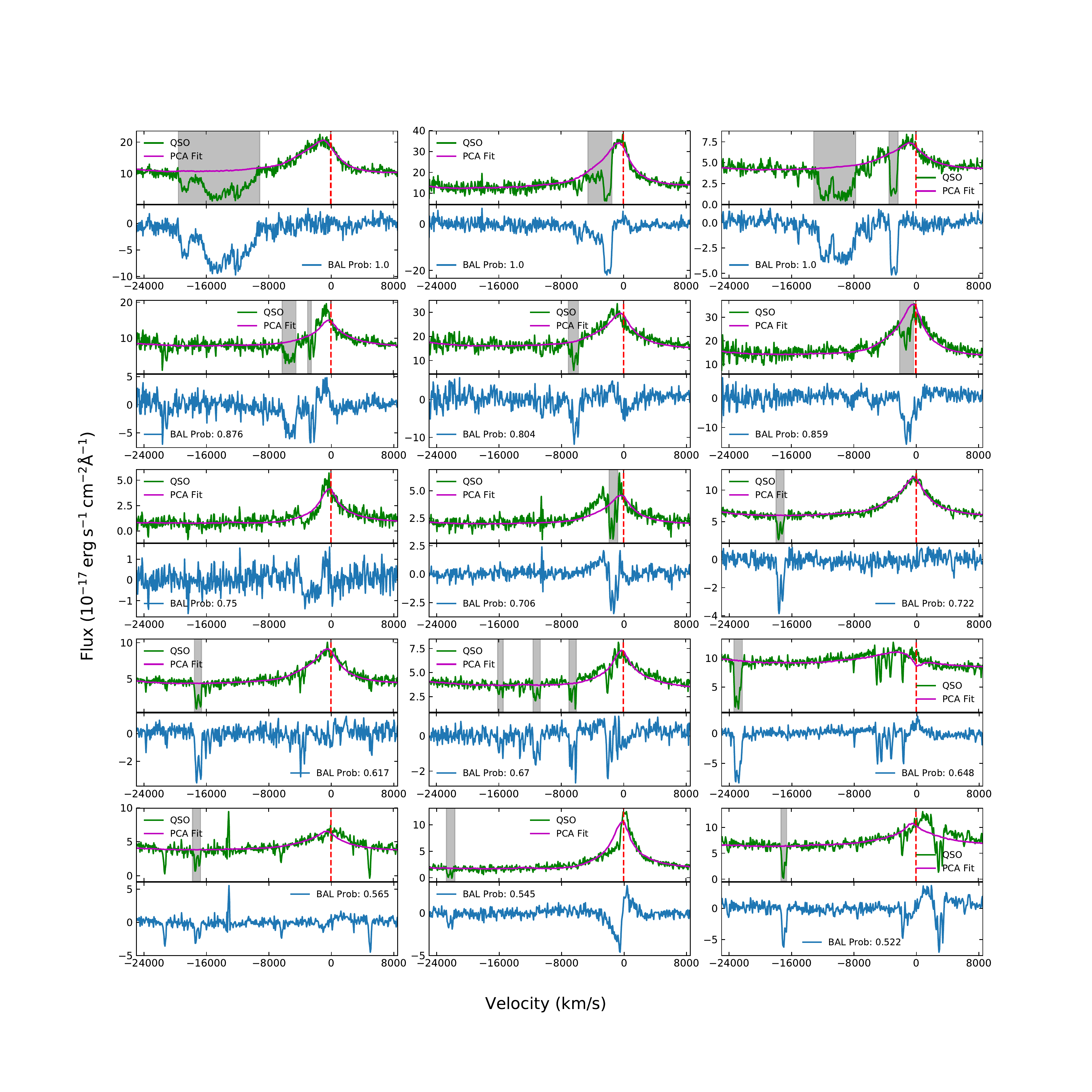}
\caption{SDSS-DR14 BAL classification examples: {\it First row:} BAL probability 90-100\%; {\it Second row:} BAL probability 80-90\%; {\it Third row:} BAL probability 70-80\%; {\it Fourth row:} BAL probability 60-70\%; {\it Fifth row:} BAL probability 50-60\%.}
\label{fig:dr14BALex}
\end{figure*}

\begin{figure*}[p]
\centering
\epsscale{1.2}
\plotone{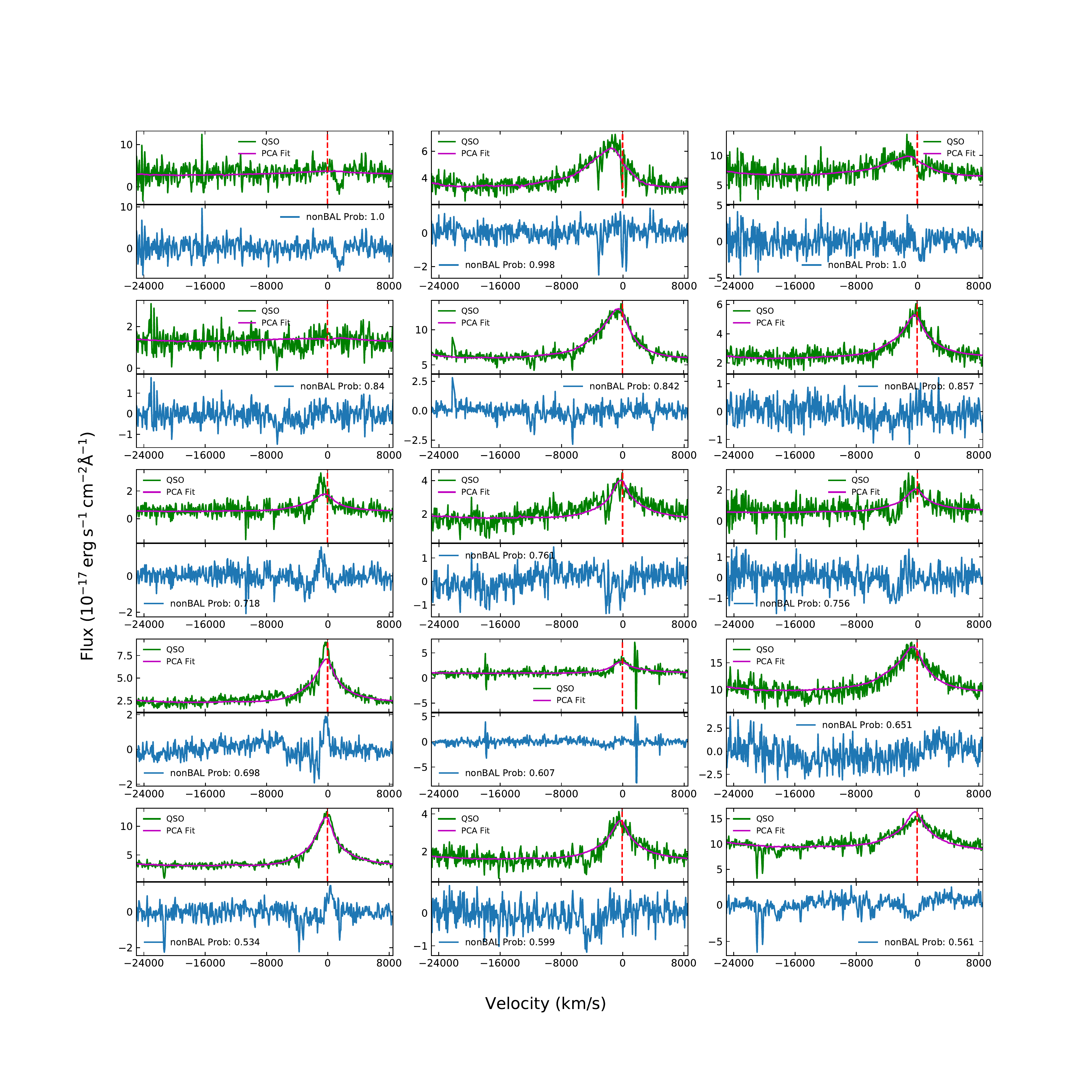}
\caption{SDSS-DR14 non-BAL classification examples: {\it First row:} non-BAL probability 90-100\%; {\it Second row:} non-BAL probability 80-90\%; {\it Third row:} non-BAL probability 70-80\%; {\it Fourth row:} non-BAL probability 60-70\%; {\it Fifth row:} non-BAL probability 50-60\%.}
\label{fig:dr14nonBALex}
\end{figure*}

Our classifier identifies 53,760 of the 320,821 quasars with a BAL probability greater than 0.5, which corresponds to 16.8\% of the quasar sample. Figure \ref{fig:prob_dist} shows the cumulative probability distribution for the BALs and non-BALs in DR14. Both distributions start at 0.5, as the BAL probability $P_{{\rm BAL}}$ is simply $P_{{\rm BAL}} = 1 - P_{{\rm non-BAL}}$, or one minus the probability it is a non-BAL quasar. Probability closer to unity corresponds to greater confidence in the classification as either a BAL or a non-BAL quasar. In Figure~\ref{fig:dr14BALex} and Figure~\ref{fig:dr14nonBALex} we present some BAL and non-BAL quasar examples that span the probability range from 50\% to 100\%. 

The cumulative probability distribution for non-BAL quasars is shallower than for the BALs. For example, the classifier assigns a non-BAL probability of 90\% or greater to over 90\% of the non-BAL sample. In contrast, the classifier assigns a BAL probability of 90\% or greater to only about 70\% of the BAL sample. The implication of this difference is that the classification of a quasar as a non-BAL has higher confidence than the classification as a BAL. There are several factors that likely contribute to this difference. One is that the PCA fits to the BAL quasars are on average poorer than the fits to non-BAL quasars, so the inputs to the classifier may have more dispersion than just what is caused by the BAL features. Another is that noise in some of the lower signal-to-noise spectra is difficult to distinguish from weak BAL features. Finally, there are simply more ways a quasar can be a BAL than not. 

\begin{deluxetable*}{clcl}[b!]
\tablecaption{DR14 BAL QUASAR CATALOG (Column 29 - Column 44) \label{tbl:dr14bal}}
\tablecolumns{4}
\tablenum{1}
\tablewidth{0pt}
\tablehead{
\colhead{Column} &
\colhead{Name} &
\colhead{Format} & 
\colhead{Description}
}
\startdata
29 & BAL\_PROB & FLOAT & BAL Probability \\
30 & TROUGH\_10K & INT32 & Flag for BAL with at least one trough with $v < -10,000 \kms$ \\
\hline
31 & BI\_CIV & DOUBLE & \ion{C}{4} Balnicity Index (BI) \\
32 & ERR\_BI\_CIV & DOUBLE & \ion{C}{4} BI uncertainty $\sigma_{BI}$\\
33 & NCIV\_2000 & INT32 & Number of troughs wider than $2000 \kms$ \\
34 & VMIN\_CIV\_2000 & DOUBLE$[5]$ & Minimum velocity of each absorption trough \\
35 & VMAX\_CIV\_2000 & DOUBLE$[5]$ & Maximum velocity of each absorption trough \\
36 & POSMIN\_CIV\_2000 & DOUBLE$[5]$ & Position of the minimum of each absorption trough \\
37 & FMIN\_CIV\_2000 & DOUBLE$[5]$ & Normalized flux density at the minimum of each absorption trough \\
38 & AI\_CIV & DOUBLE & Absorption Index (AI) \\
39 & ERR\_AI\_CIV & DOUBLE & AI uncertainty $\sigma_{AI}$ \\
40 & NCIV\_450 & INT32 & Number of troughs wider than $450 \kms$ \\
41 & VMIN\_CIV\_450 & DOUBLE$[17]$ & Minimum velocity of each absorption trough \\
42 & VMAX\_CIV\_450 & DOUBLE$[17]$ & Maximum velocity of each absorption trough \\
43 & POSMIN\_CIV\_450 & DOUBLE$[17]$ & Position of the minimum of each absorption trough \\
44 & FMIN\_CIV\_450 & DOUBLE$[17]$ & Normalized flux density at the minimum of each absorption trough \\
45 & BI\_SIV & DOUBLE & BI in the \ion{Si}{4} region\\
46 & ERR\_BI\_SIV & DOUBLE & BI uncertainty in \ion{Si}{4} region\\
47 & NSIV\_2000 & INT32 & Number of \ion{Si}{4} troughs wider than $2000 \kms$ \\
48 & VMIN\_SIV\_2000 & DOUBLE$[5]$ & Minimum velocity of each absorption trough \\
49 & VMAX\_SIV\_2000 & DOUBLE$[5]$ & Maximum velocity of each absorption trough\\
50 & POSMIN\_SIV\_2000 & DOUBLE$[5]$ & Position of the minimum of each absorption trough \\
51 & FMIN\_SIV\_2000 & DOUBLE$[5]$ & Normalized flux density at the minimum of each absorption trough \\
52 & AI\_SIV & DOUBLE & Absorption index (AI) in \ion{Si}{4} region \\
53 & ERR\_AI\_SIV & DOUBLE & AI uncertainty in \ion{Si}{4} region \\
54 & NSIV\_450 & INT32 & Number of absorption trough wider than 450 $km s^{-1}$ \\
55 & VMIN\_SIV\_450 & DOUBLE$[17]$ & Minimum velocity of each  absorption trough \\
56 & VMAX\_SIV\_450 & DOUBLE$[17]$ & Maximum velocity of each  absorption trough\\
57 & POSMIN\_SIV\_450 & DOUBLE$[17]$ & Position of the minimum of each absorption trough \\
58 & FMIN\_SIV\_450 & DOUBLE$[17]$ & Normalized flux density at the minimum of each absorption trough \\
\enddata
\end{deluxetable*}

Our SDSS DR14 BAL catalog follows a similar format to the DR12 BAL catalog by \citet{Paris17}, and the first 28 columns have identical information. The contents of the remaining columns are described in Table~\ref{tbl:dr14bal}. These include the BAL probability from our CNN classifier, AI and BI, the number of troughs, blueshift and velocity range of each trough, and the normalized flux density at the minimum of each trough. 

\begin{figure*} 
\centering
\epsscale{1.15}
\plottwo{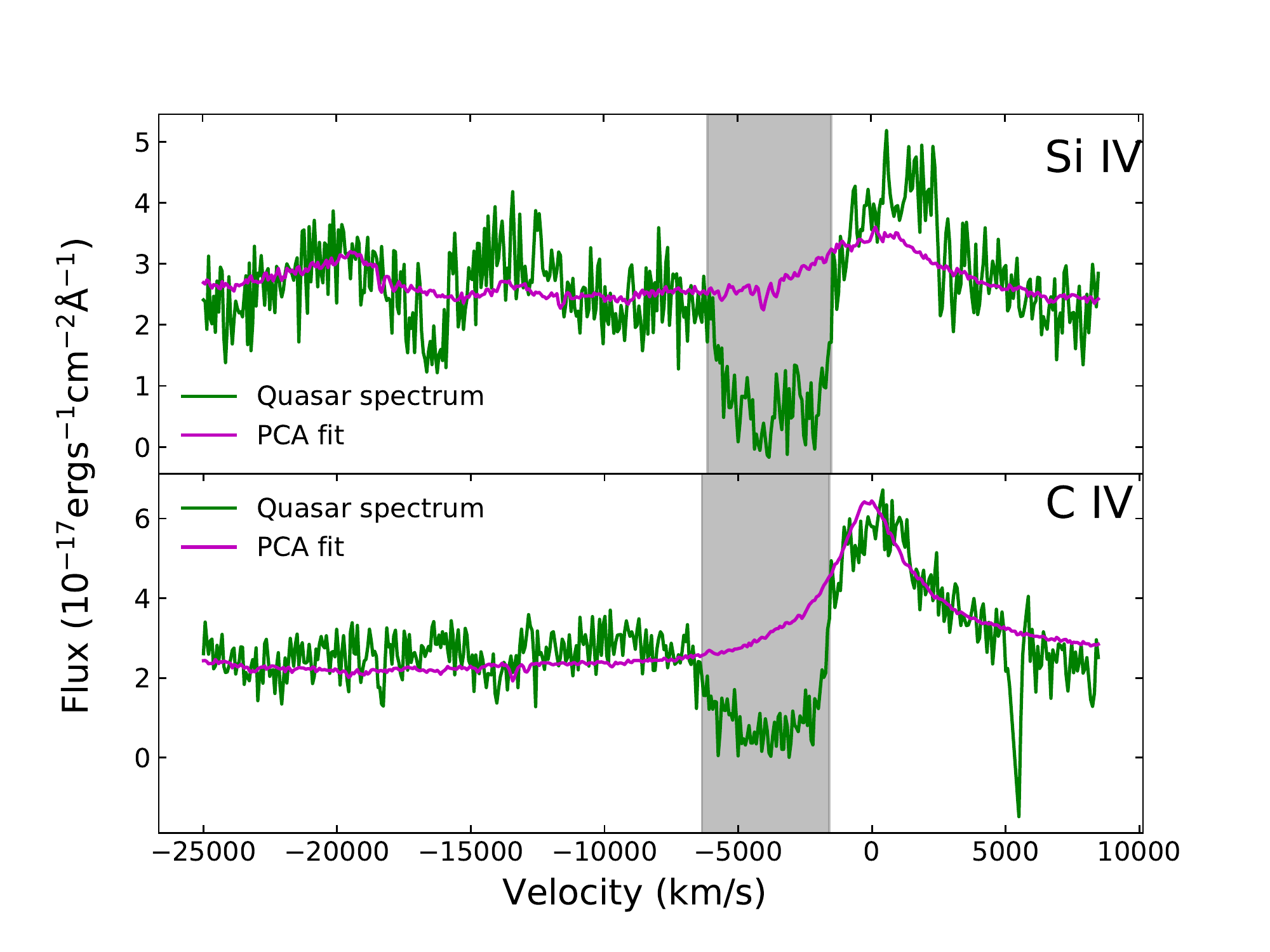}{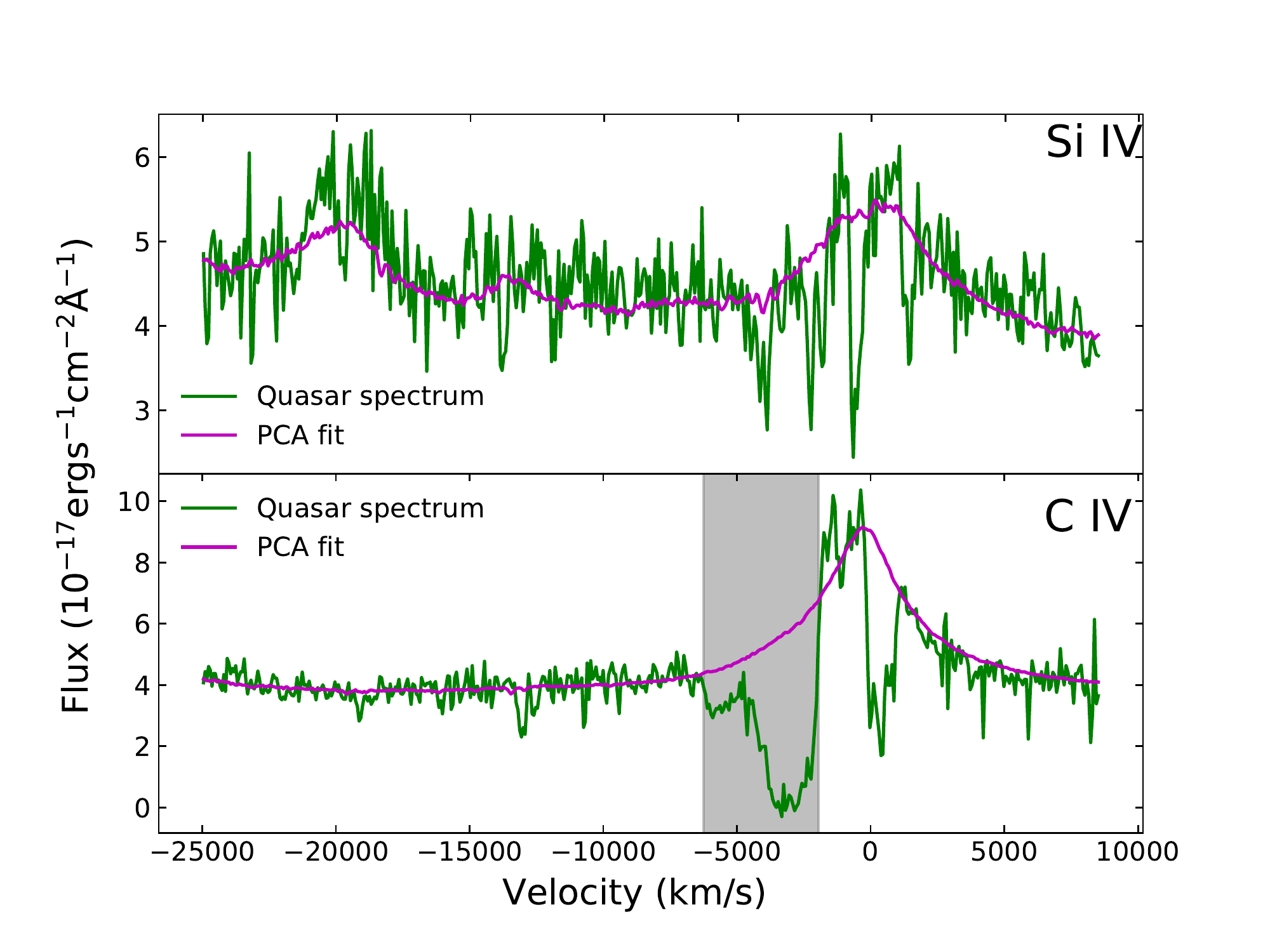}
\caption{Examples of our search for \ion{Si}{4} troughs. {\it Left panels:} Quasar with BAL features present in both \ion{C}{4} and \ion{Si}{4}. {\it Right panels:} Quasar with BAL features only present in \ion{C}{4}.}
\label{fig:SiIV}
\end{figure*}

We also present these quantities for troughs associated with the \ion{Si}{4} region. The ionization potential of \ion{Si}{4} is sufficiently different from \ion{C}{4} that the two lines may be used together to probe the ionization level, kinematics, and column density of the outflow \citep[e.g.][]{FilizAk14,Baskin15}. Approximately half (48\%) of the BALs have significant AI or BI values for the \ion{Si}{4} region, and specificially 54\% (40\%) of those with a significant AI (BI) value in \ion{C}{4} have a significant AI (BI) in \ion{Si}{4}. Figure~\ref{fig:SiIV} shows example absorption troughs for \ion{Si}{4} and \ion{C}{4} for two quasars.

\begin{figure*}[t] 
\centering
\plottwo{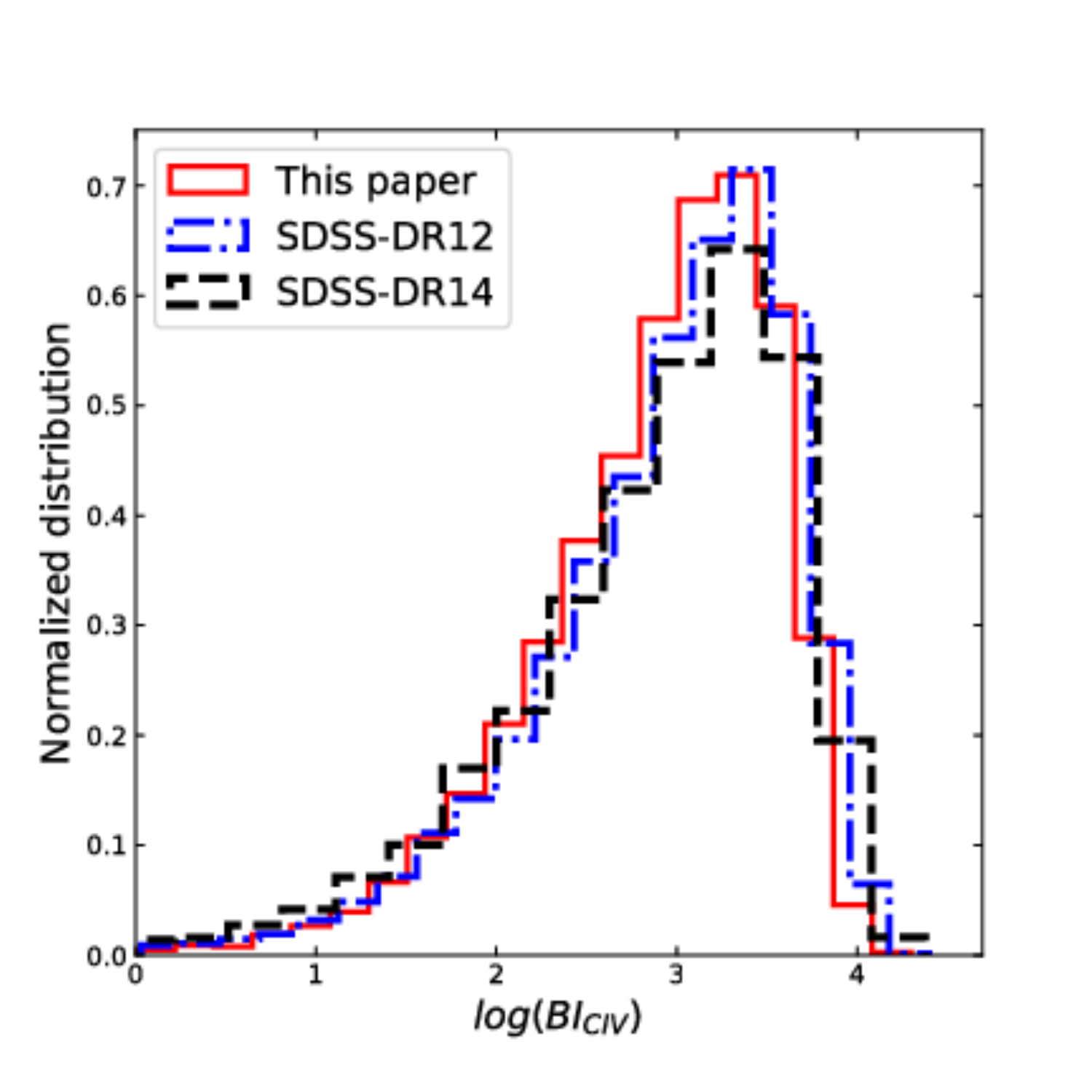}{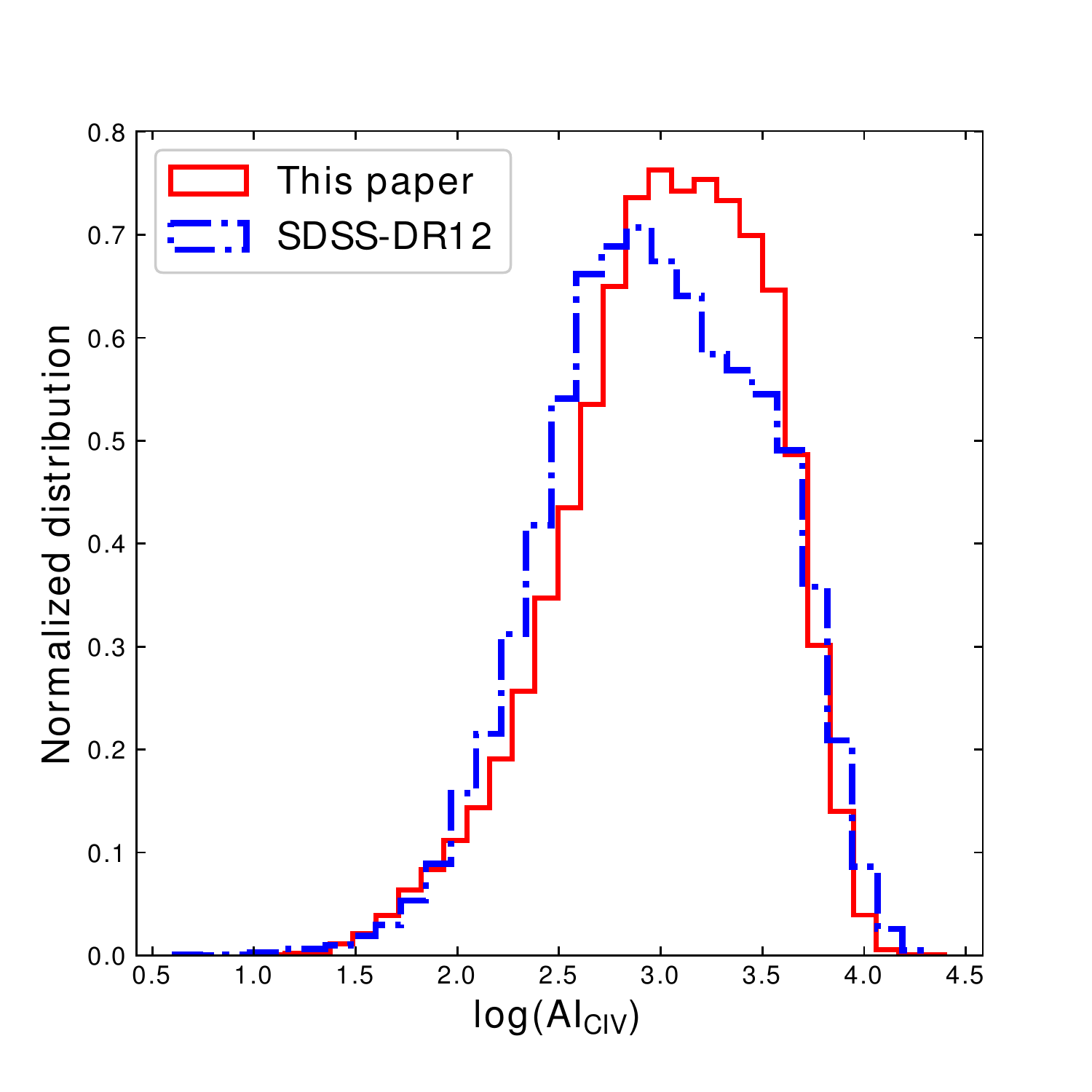}
\caption{{\it Left panel:} Balnicity index (BI) distribution for DR14 BAL quasars calculated in our new catalog ({\it solid red}), BI calculated by \citet{Paris18} ({\it dashed blue}), and DR12 BI value in \citet{Paris17} ({\it dashed black}). {\it Right panel:} Absorption index (AI) distribution for the new DR14 BAL quasars catalog and DR12 AI value in \citet{Paris17}.}
\label{fig:AI_BI_Dist}
\end{figure*}

\section{Comparison to Previous Catalogs} \label{sec:compare}

In this section we compare our classifications and measurements to results presented as part of the SDSS DR12 \citep{Paris17} and DR14 \citep{Paris18} quasar catalogs. This includes both the classification of quasars as BALs and quantities commonly used to characterize the properties of the BAL troughs. 

\subsection{Comparison to DR12} \label{sec:diff1214}

The SDSS DR12 catalog includes a visual BAL classification, along with measurements of many quantities that are commonly used to parameterize BAL features. These include the quantities BI, AI, $\chi^2_{trough}$, and the velocities and velocity widths of the absorption systems. As all of the quasars in the DR12 catalog are included in DR14, we used the DR12 catalog to both quantify the performance of our CNN classifier and check our measurements of the same BAL quantities.

There are numerous criteria that could be used to identify BAL quasars. Our classifier was trained based on our own visual criteria as described in Section~\ref{sec:train} and assigns $P_{BAL} > 50$\% to 38,653 of the DR12 quasars. We compared our classifier to the DR12 visual flag and recover 93\% of those BALs. This difference is not surprising, as we reclassified approximately 6.5\% of the DR12 quasars in our training set. Two other ways to more quantitatively identify BALs are to require a significant AI measurement, which we define as AI $> 3\sigma_{AI}$, and a significant BI measurement. Our classifier labels 96.0\% and 98.6\% of the DR12 quasars with significant AI and BI, respectively, as BAL quasars. The greater agreement for significant BI relative to significant AI is because the equivalent width of the absorption is almost always greater for a quasar with a significant BI measurement. Lastly, we also examined the performance of the classifier on the subset of particularly high-velocity outflows with $v < -10,000 \kms$. The motivation for this choice is that \ion{N}{5} systems at these velocities will overlap the wavelength range commonly used for studies with the Ly-$\alpha$ forest. Our classifier identifies 95.1\% of the DR12 quasars with BAL troughs at $v < -10,000 \kms$. 

Recently \citet{Busca18} presented a CNN quasar classifier called QuasarNET. Their classifier was designed and trained to identify quasars in spectra of quasar targets (to distinguish quasars from stars and emission line galaxies) and determine redshifts. Their code includes a feature detection unit specific to broad absorption associated with \ion{C}{4}, in addition to units for prominent emission lines that were used for the redshift measurement. \citet{Busca18}'s QuasarNET applies a BAL classification to $98.0 \pm 0.4$\% of quasars with a visual BAL classification and BI\_CIV$>0$ from DR12. There are 15,1777 quasars in DR12 that meet both of these criteria. Similarly, it applies a non-BAL classification to $97.0 \pm 0.2$\% of the quasars visually classified as non-BALs and that have BI\_CIV$=0$ from DR12. For these exact same criteria, our classifier identifies $97.4$\% of BALs in DR12 with both visual classifications and BI\_CIV$>0$ from DR12, and identifies $94$\% of the corresponding sample of non-BALs. These values are in quite good agreement, given the differences in criteria used to construct the training sets for their and our classifiers. 

\citet{Busca18} also investigate the completeness and purity of the BAL classification. Completeness is the fraction of all true BALs that the classifier identifies correctly, and the purity is the fraction of true BALs in the sample that the classifier identifies as BALs. Typically classification schemes optimize one of these quantities at the expense of the other. For cosmological applications, the completeness is more important in order to account for the impact of BALs on the analysis, and BALs are a small fraction of the quasar sample. The completeness and purity of the BALs in DR12 are not exactly known, so instead \citet{Busca18} define and measure the ``pseudo-purity" and ``pseudo-completeness" of the sample produced by their classifier. The definitions of both quantities use those BALs in DR12 with both a visual BAL classification and BI\_CIV$>0$ to define the population of ``true BALs." The pseudo-purity is then the number of BALs identified by their classifier that also have either BI\_CIV$>0$ or a visual BAL classification in DR12 divided by the total number of BALs identified by their classifier. They define the pseudo-completeness as the number of ``true BALs" identified by their classifier divided by the total number of ``true BALs." Note that the pseudo-completeness is defined the same way as the 98\% classification in the previous paragraph. \citet{Busca18} measure $98.0 \pm 0.4$\% for the pseudo-completeness and $77\pm1$\% for the pseudo-purity. For these exact same criteria, we measure $97.4$\% and $40$\% for the pseudo-completeness and pseudo-purity, respectively. Our classifier does comparably well in terms of completeness to \citet{Busca18}, which is more relevant to cosmological analysis. The pseudo-purity is much lower. This is because our threshold for a BAL is much more inclusive than the requirement of both visual BAL classification and BI\_CIV$>0$ adopted by \citet{Busca18}. For example, 93\% of the quasars we classify as BALs have a significant AI value. 

We also compare our measurements of AI, BI, and BAL velocities with the values reported in \citet{Paris17}. Figure~\ref{fig:AI_BI_Dist} shows the normalized distribution of AI and BI values for our values and the DR12 and DR14 values. The BI distributions are similar, while our catalog has more AI values with $log\,{AI}_{CIV}$ between 3 and 3.5. The discrepancies are likely caused by differences in the continuum fits between those used by \citet{Paris17} and our work. Differences in the continuum have a bigger impact on AI than BI because in general the fits are more uncertain near the \ion{C}{4} line, and AI values can extend to the systemic redshift. All of our measurements also use the DR14 spectroscopic data release, which included re-reductions of BOSS spectra with the eBOSS pipeline \citep{Abolfathi18}.

\subsection{Comparison to DR14}

\begin{figure*}[t]
\includegraphics[scale = 0.4]{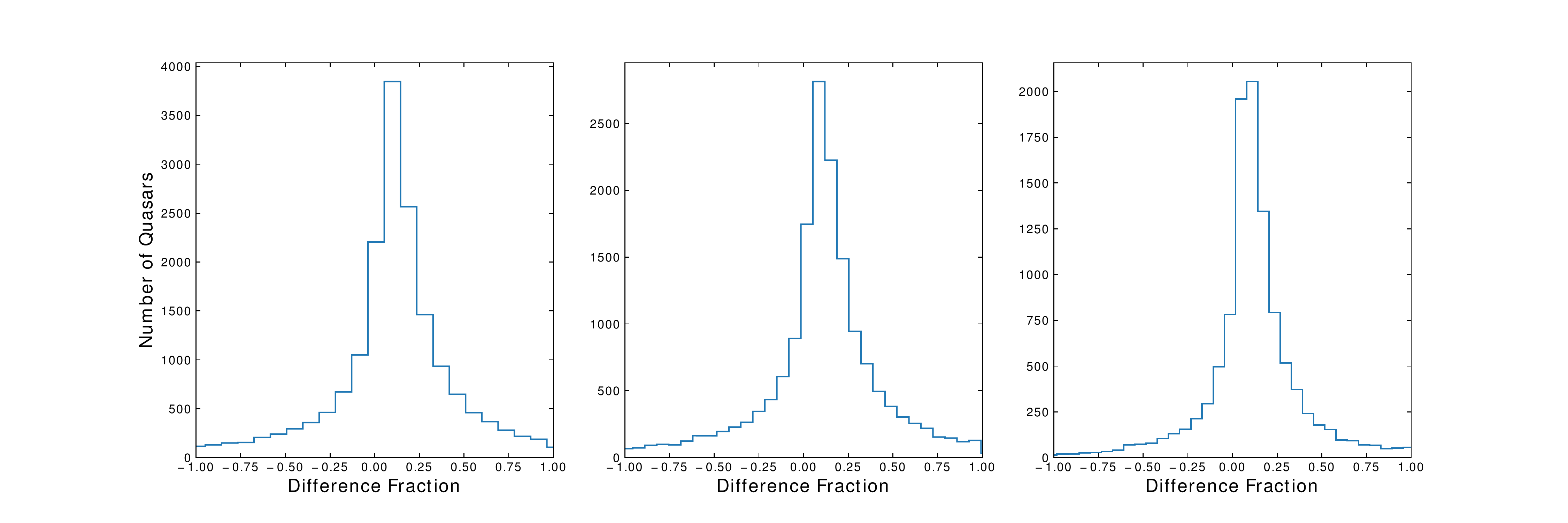}
\centering
\caption{The fractional difference in BI between this paper and the DR14 quasar catalog of \citet{Paris18} for three different sub-sets of BAL quasars in DR14: quasars with $BI > 0 \kms$ ({\it left}); $BI > 3\sigma_{BI}$ ({\it middle}); $BI > 1000 \kms$ ({\it right}).}
\label{fig:BI DIFF DR14}
\end{figure*}

The DR14 quasar catalog includes BI and its uncertainty $\sigma_{BI}$, but not 
a visual classification flag or other BAL properties. Figure~\ref{fig:AI_BI_Dist} shows the BI distribution from DR14, along with the BI distributions from our measurements and the DR12 catalog. The BI distributions for all three catalogs are in good agreement. The DR14 catalog BI values from \citet{Paris18} are based on the same parent sample of quasars, although in some cases our BI measurements are different from those in the DR14 catalog for the same quasars. 

We illustrate the agreement between our BI values and those in the DR14 catalog with the fractional difference: 
\begin{equation}
Frac = \frac{BI - BI_{P18}} {BI_{P18}}.
\end{equation}
The quantity BI is our measurement and BI$_{P18}$ is the value from \citet{Paris18}, and we compute the fractional difference for every quasar where both we and \citet{Paris18} report a BI value. Figure~\ref{fig:BI DIFF DR14} shows the distribution of this fractional difference for three subsamples of quasars: all quasars with $BI > 0$, all quasars with significant BI values, and all quasars with $BI > 1000$. Generally the values agree within 25\%, and we attribute these differences to the continuum fits.

\section{Summary}

We developed and trained a Convolutional Neural Network to automatically classify BAL quasars and applied it to spectroscopy of the quasars in the SDSS DR14 quasar catalog of \citet{Paris18}. We classify 53,760 out of 320,821 quasars with $1.57 < z < 5.56$ as BALs, or 16.8\% of the sample. These are quasars with a BAL probability of at least 50\% according to our classifier. We trained our classifier to be sensitive to even relatively shallow BAL troughs and absorption features from a blueshift of $-25,000 \kms$ to the center of the \ion{C}{4} emission line. We have demonstrated that our BAL classifications and measurements of various BAL parameters such as AI and BI are in good agreement with previous work.

We have also measured the BAL properties of all of these quasars. These properties include the AI and BI values for each quasar, as well as the blueshifts of the troughs, and their velocity width. Our catalog contains these values for absorption associated with both the \ion{C}{4} and \ion{Si}{4} lines. One potential application of this larger BAL catalog is to identify rare subsets of BAL quasars for detailed analysis. Another is to stack multiple, similar BAL spectra to create higher signal-to-noise ratio spectra, as has been done for Dampled Lyman $\alpha$ systems \citep{MasRibas17} and previous BAL samples \citep{Hamann19}. Such spectra could be combined with spectral synthesis codes, such as {\it simBAL} \citep{Leighly18}, for more detailed studies of a wide variety of BAL subclasses. 

The identification of BALs in this quasar catalog will improve the value of these quasars for cosmological analysis, as well as to identify rare subsets of BAL quasars for studies of AGN physics. Two cosmological applications are to improve redshifts in the case where absorption features compromise one or more of the strong emission lines, and to account for potential contamination of the Ly$\alpha$ forest absorption by \ion{N}{5}, \ion{O}{6}, and perhaps \ion{P}{5} and Ly$\alpha$. The strengths and velocities of \ion{C}{4} and \ion{Si}{4} troughs may be useful to predict the strengths and locations of other features that may contaminate the Ly$\alpha$ forest. This CNN classifier can also be readily adopted to larger spectroscopic samples of quasars from the upcoming DESI \citep{Desi16a,Desi16b} and 4MOST \citep{deJong14} surveys. 

\acknowledgements
We are grateful to Karen Leighly, Don Terndrup, and Lluis Mas Ribas for helpful discussions. We also thank Gerard Lemson, Ani Thakar, and Joel Brownstein for support with SciServer. This material is based upon work supported by the U.S. Department of Energy, Office of Science, Office of High Energy Physics under Award Number DE-SC-0011726.

SciServer is a collaborative research environment for large-scale data-driven science. It is developed at, and administered by, the Institute for Data Intensive Engineering and Science at Johns Hopkins University. SciServer is funded by the National Science Foundation Award ACI-1261715. For more information about SciServer, please visit http://www.sciserver.org.

Funding for the Sloan Digital Sky Survey IV has been provided by the Alfred P. Sloan Foundation, the U.S. Department of Energy Office of Science, and the Participating Institutions. SDSS-IV acknowledges support and resources from the Center for High-Performance Computing at the University of Utah. The SDSS web site is www.sdss.org.

SDSS-IV is managed by the Astrophysical Research Consortium for the  Participating Institutions of the SDSS Collaboration including the Brazilian Participation Group, the Carnegie Institution for Science, Carnegie Mellon University, the Chilean Participation Group, the French Participation Group, Harvard-Smithsonian Center for Astrophysics, Instituto de Astrof\'isica de Canarias, The Johns Hopkins University, Kavli Institute for the Physics and Mathematics of the Universe (IPMU) / University of Tokyo, the Korean Participation Group, Lawrence Berkeley National Laboratory, Leibniz Institut f\"ur Astrophysik Potsdam (AIP),  Max-Planck-Institut f\"ur Astronomie (MPIA Heidelberg), Max-Planck-Institut f\"ur Astrophysik (MPA Garching), Max-Planck-Institut f\"ur Extraterrestrische Physik (MPE), National Astronomical Observatories of China, New Mexico State University, New York University, University of Notre Dame, Observat\'ario Nacional / MCTI, The Ohio State University, Pennsylvania State University, Shanghai Astronomical Observatory,  United Kingdom Participation Group, Universidad Nacional Aut\'onoma de M\'exico, University of Arizona, University of Colorado Boulder, University of Oxford, University of Portsmouth, University of Utah, University of Virginia, University of Washington, University of Wisconsin, Vanderbilt University, and Yale University.

\end{document}